\documentclass{article}

\usepackage{arxiv}

\usepackage[utf8]{inputenc} % allow utf-8 input
\usepackage[T1]{fontenc}    % use 8-bit T1 fonts
\usepackage{hyperref}       % hyperlinks
\usepackage{url}            % simple URL typesetting
\usepackage{booktabs}       % professional-quality tables
\usepackage{amsfonts}       % blackboard math symbols
\usepackage{nicefrac}       % compact symbols for 1/2, etc.
\usepackage{microtype}      % microtypography
\usepackage{lipsum}
\usepackage{graphicx}
\usepackage{fancyhdr}
\usepackage[most]{tcolorbox}
\usepackage{pgfplots}
\pgfplotsset{compat=1.18}
\usepackage{longtable}
\graphicspath{ {./images/} }

\newcommand{\password}{\textit{\texttt{pw}}}
\newcommand{\fingerprint}{\textit{\texttt{fp}}}

\newcommand{\iris}{\textit{\texttt{ir}}}
\newcommand{\pin}{\textit{\texttt{pin}}}
\newcommand{\otp}{\textit{\texttt{otp}}}

\newcommand{\aqua}{\textsc{Aqua}}
\newcommand{\aquaa}{\textsc{Aqua-A}}
\newcommand{\aquat}{\textsc{Aqua-T}}
\newcommand{\aquas}{\textsc{Aqua-S}}

\title{AQUA - A Modular Authentication Quality Modeling Framework}

\author{
  Alex R. Mattukat \\
  Research Group Software Construction\\
  RWTH Aachen University\\
  Ahornstraße 55 \\
  Aachen, Germany \\
  \texttt{mattukat@swc.rwth-aachen.de}
  \And
  Michael Zerbe \\
  Research Group Software Construction\\
  RWTH Aachen University\\
  Ahornstraße 55 \\
  Aachen, Germany \\
  \texttt{michael.zerbe@rwth-aachen.de} \\
  \And
  Krisa Carka \\
  RWTH Aachen University\\
  Ahornstraße 55 \\
  Aachen, Germany \\
  \texttt{krisa.carka@rwth-aachen.de} \\
}

\begin{document}
\tiny
This is the \textbf{peer-reviewed, accepted version of a paper}. It will appear in the proceedings of the 13\textsuperscript{th} International Workshop on Quantitative Approaches to Software Quality (\textbf{QuASoQ 2026}), co-located with the 33\textsuperscript{rd} Asia-Pacific Software Engineering Conference (\textbf{APSEC 2026}). The final published version will be available in the CEUR Workshop Proceedings (CEUR-WS.org, ISSN 1613-0073). © 2026 for this paper by its authors. Use permitted under Creative Commons License Attribution 4.0 International (CC BY 4.0).
\normalsize

\maketitle
\thispagestyle{fancy}

\begin{abstract}
Authentication solutions are the key design, architectural, and implementation decisions that implement authentication security controls. An example is a password-based authentication solution. While researchers often examine authentication quality, existing approaches do not distinguish which effects on quality stem from which aspect of authentication. For example, multi-factor authentication solutions tend to be less usable because they combine multiple authenticators, whereas passkeys are more secure than passwords because of properties inherent to the authenticator itself. This paper presents \aqua, a modular quality modeling framework for authentication solutions. \aqua’s core idea is to decompose authentication into four interrelated concepts, enabling modular specification of authentication quality and tracing quality effects back to their source. We demonstrate \aqua\ with initial definitional quality models that address the levels of authenticators, authentication techniques, and authentication solutions. We further illustrate the interrelations between these concepts through concrete examples and a usability metric, showing how quality propagates across the concepts and ultimately affects the overall quality of an authentication solution. Our results provide a foundation for future work, including designing assessment and prediction models for authentication solutions and automated quality assessments.
\end{abstract}

\newtcolorbox{mainBox}[1]{%
		enhanced,
		colback=white!90!black,
		colframe=darkgray,
		colbacktitle=darkgray,
		coltitle=white,
		fonttitle=\bfseries,
		title=#1,
		rounded corners,
		boxrule=3pt,
		drop shadow=darkgray,
		width=\textwidth,
	}

% keywords can be removed
  
\section{Introduction}\label{sec:introduction}

Security controls are central to securing software systems. They are countermeasures that protect a software system's confidentiality, integrity, and availability \cite{NIST.800-37}. Examples are authentication and Distributed Denial-of-Service (DDoS) prevention. The key design, architectural, and implementation choices made to realize a security control are known as a security solution \cite{NIST.800-172}.

Security solutions exhibit quality properties, such as usability and performance. However, they vary significantly depending on the security control they implement. For example, security solutions for authentication focus on how users can securely authenticate within a system~\cite{Todorov.2007}; security solutions for DDoS prevention focus on load balancing and blocking suspicious traffic~\cite{Mahjabin.2017}. Comparing the usability of an authentication solution with that of a DDoS-prevention solution makes little sense, as they depend on entirely different factors. To elaborate on the relationship between quality and security solutions, we thus need to focus on a specific class of security solutions. We focus on authentication, as it is one of the most widely used security controls \cite{NIST.800-63-4}. Moreover, authentication failures rank 7th on the OWASP\footnote{Open Worldwide Application Security Project} Top Ten list~\cite{OWASP.Top10}, the de facto standard for security risks in web applications, underlining the need to understand what determines the quality of authentication solutions.

\paragraph{Background:}
Authentication is the process by which a subject---in the following referred to as the \textit{user}---gives proof of the identity they claim using one or more authenticators~\cite{Todorov.2007}. An \textit{authenticator} is the entity that a user controls and possesses to prove their identity, such as a password or fingerprint~\cite{NIST.800-63-4}. We refer to the key design, architectural, and implementation choices made to realize authentication as an \textit{authentication solution}. 

Authentication solutions, being security solutions, exhibit quality properties. For example, password-based authentication is less secure than passkey-based authentication~\cite{Matzen.2025}, and multi-factor authentication (MFA) solutions are more secure but tend to be less usable than single-factor authentication solutions~\cite{Klivan.2023}. While such comparisons may seem obvious (MFA requires at least two authenticators, which implies a negative effect on usability), the root causes of these qualitative characteristics are less obvious. Some effects stem from using multiple authenticators, others from inherent properties of the authenticator (a passkey is more secure than a password because it prevents users from choosing weak passwords), and still others from the technical implementation (a solution that enforces a strong password is more secure than one without password-strength policies).

\paragraph{Problem Statement:}
Researchers have studied the relationship between authentication and quality extensively, but they use inconsistent terminology, a well-known problem in the security field~\cite{Neil.2023}. For example, some examine the quality of authentication \emph{schemes}~\cite{Bonneau.2012}, others that of authentication \emph{methods}~\cite{Furnell.2022} or \emph{mechanisms}~\cite{Chenchev.2021}, and still others that of authentication in general~\cite{Zafar.2015}. It remains unclear whether these refer to the same quality, since the differences between an authentication scheme, method, and mechanism, and how each relates to authentication in general, are not defined. This ambiguity makes results hard to compare across studies.

A second, deeper problem is that existing models do not distinguish whether a quality effect stems from the authenticator, from the way authenticators are employed, or from the concrete implementation. This makes it impractical to trace how quality propagates from the authenticators through their employment to the solution as a whole, and thus to explain why a solution has the quality it has. For example, how does the usability of an MFA solution that uses a fingerprint and a hardware one-time-password (OTP) token compare with one that uses a short PIN and voice recognition? Answering this requires knowing which effects come from each authenticator, which come from combining them, and which come from the technical realization. To assess such nuances, we need tailored quality models that capture the
different facets of what makes an authentication solution and how each facet affects its quality. Such models could help architects make well-informed decisions, compare solution alternatives by their estimated quality, and allow automated quality assessment of authentication solutions in a system. To our knowledge, no such quality model exists. 

\paragraph{Contribution:} 

To close this gap, we developed \aqua, a modular quality modeling framework for authentication solutions. \aqua\ decomposes authentication into four interrelated concepts: the \emph{authenticator} (the entity a user possesses or controls to perform authentication, e.g., a password), the \emph{authenticator employment} (a composition scheme for authenticators), the \emph{authentication technique} (the concrete composition of authenticators under an authenticator employment), and the \emph{authentication solution} (the concrete design, architecture, and implementation that realizes a technique). This decomposition separates the three sources of quality effects identified above: the properties inherent to an authenticator, the effects of how authenticators are combined, and the effects of the concrete implementation.

Therefore, our key contribution is not another security quality model, but a quality modeling framework that enables the development of tailored, modular quality models for authentication solutions by ensuring that the conceptual separation of key authentication concepts is maintained. It clearly distinguishes itself from related quality evaluation and modeling approaches~\cite{Bonneau.2012,Furnell.2022,Zafar.2015}, as no approach we know of accounts for this conceptual distinction. %We discuss this in more detail in Section~\ref{sec:related-work}.

We demonstrate \aqua\ through three initial quality models: \aquaa\ for authenticators, \aquat\ for authentication techniques, and \aquas\ for authentication solutions. Moreover, we show how quality propagates from inherent authenticator properties, through their use in a technique, to their realization in a solution, using an example usability metric. The following research question underlies this study:

\vspace{5pt}

\noindent \textbf{RQ}: How can a quality modeling framework for authentication solutions be designed that captures how quality propagates from the authenticators through their employment to the solution?

\subsection{Methodology \& Outline}
This is a Design Science Research (DSR) study \cite{Peffers.2007}, as we develop a research artifact (\textsc{Aqua}). The six DSR steps are mapped to our study as follows: (1) The problem was identified, and a solution was motivated as presented above. (2) The research objective is reflected by the research question. (3) We developed, verified, and refined the Aqua conceptual model step by step. To verify the model’s internal consistency and expressiveness, we formalized the concepts. This confirmed that the model can represent single-factor, multi-factor, and adaptive authentication, including nested compositions. (4) \aqua's demonstration has two parts: First, we demonstrate each concept and how it relates to quality with comprehensive examples, including metrics. Second, we present three quality models for these concepts that we developed with \aqua. (5) Due to the early stage of this project, a comprehensive evaluation is missing; we address it in future work. (6) Communication is done via scientific publications.

Section \ref{sec:related-work} presents related work. In Section \ref{sec:quality_of_authentication}, the conceptual decomposition of authentication on which the \textsc{Aqua} framework is built is introduced. Next, Section \ref{sec:quality} presents their relationship to quality, including demonstration examples. In Section~\ref{sec:demo} we demonstrate \textsc{Aqua} using initial quality models we built with it. Section \ref{sec:conclusions} concludes this paper.

\section{Related Work} \label{sec:related-work}

Quality models are established for modeling security quality by hierarchically structuring security into more detailed factors and subfactors (e.g., \cite{Zafar.2015,Xu.2013,Aziz.2018}). 
Some studies also address specific quality factors in the context of authentication. Bonneau et al. \cite{Bonneau.2012} propose a comparison framework of web authentication schemes based on usability, deployability, and security. Furnell et al. \cite{Furnell.2022} address the accessibility of authentication methods. Both studies provide important insights into the specificity of authentication-related quality as well as into possible approaches for evaluating these quality factors. Our approach differs from all approaches we know of because it holistically models authentication quality through a modular decomposition of authentication concepts. This makes it possible to trace the effects of various design decisions on quality back to the conceptual components of authentication.

\textsc{Aqua} builds on our prior work. In~\cite{Authentication.Techniques}, we identified the relationships between the concepts on which \aqua\ is based and showed that these can be treated as distinct first-order concepts. In that paper, we also identified the quality dimension as a limitation and a direction for future work. In~\cite{Authentication.Patterns} we proposed an authentication pattern catalog. Patterns in software engineering encapsulate qualitative aspects through their forces and consequences. \aqua\ incorporates our insights from both papers and synthesizes them into a holistic quality modeling framework for authentication.

%In \cite{Authentication.Techniques} we developed initial concepts 

\section{The \textsc{Aqua} Framework}\label{sec:quality_of_authentication}
\begin{figure}[thb]
    \begin{center}
        \includegraphics[width=0.9\textwidth]{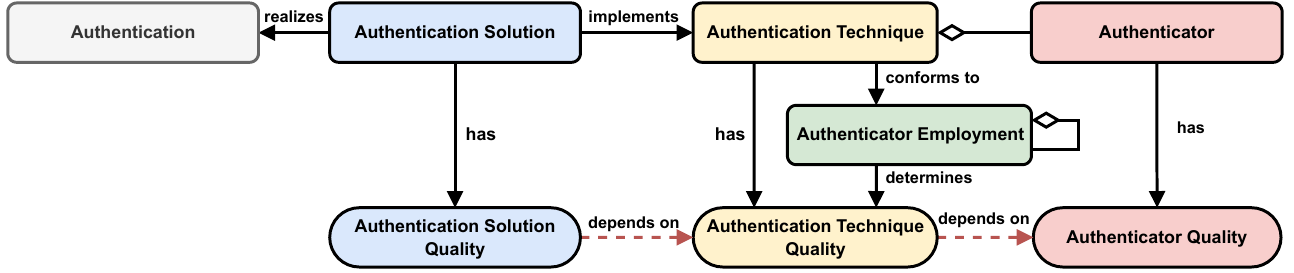}
    \caption{The core concepts and their relationships that make up the \textsc{Aqua} framework.}
    \label{fig:concepts}
    \end{center}
\end{figure}

As introduced above, \textsc{Aqua} decomposes authentication into four interrelated concepts. This modular decomposition enables flexible, adaptable authentication solutions based on the authenticators and how they are used. This section presents the decomposition. Figure~\ref{fig:concepts} depicts \textsc{Aqua}'s concept model.

\subsection{Authenticator}\label{subsec:authenticator}
An authenticator represents the most fundamental part of authentication. It is the entity a user possesses, controls, and uses to perform authentication. In \aqua, an authenticator denotes the type level, e.g., a password or a fingerprint; its instances are concrete authenticators.

\paragraph{Examples:}
\begin{enumerate}
    \item A password (\password), fingerprint (\fingerprint), iris (\iris), PIN (\pin) and one-time password (\otp) are authenticators. These authenticator abbreviations will be used in the remainder of the paper.
    \item Alice's \password\ ``PWD123!'' and Bob's \fingerprint\ are concrete authenticators.
\end{enumerate}

\subsection{Authenticator Employment}

Many authentication solutions do not use a single authenticator, but compose several authenticators together. The way they are composed matters independently of which authenticators are used. For example, ``first use a \password\, then use a \fingerprint'' and ``first scan the \iris, then use a \pin'' reveal the same structural feature: the sequential use of two different authenticators. 

To capture this structural composition, we define the \emph{authenticator employment}. An authenticator employment is a composition scheme for authenticators. To stay independent of concrete authenticators, it is built on ``slots''. A slot is a place that awaits exactly one authenticator. An employment specifies how many slots there are (\emph{cardinality}) and how they are used together (\emph{mode} and \emph{order}). The authentication technique, defined in Section~\ref{sec:technique}, determines which authenticator goes into which slot.

An employment's cardinality is the number of slots it specifies; its order is the position of each slot from left to right. The simplest employment specifies exactly one slot, i.e., has a cardinality of one. We refer to it as \emph{single employment}. A single employment models single-factor authentication~\cite{Ometov.2018}. Employments with cardinality $>1$ are \emph{multi employments}. In addition, they define a mode. The mode specifies how the authenticators assigned to its employment slots are used together.

We distinguish between three modes that multi employments can specify:
\begin{itemize}
  \item \textbf{Sequential employment} ($E_{seq}$): All slots are used, one after the other, in a fixed order. Common MFA solutions use this mode~\cite{Ometov.2018}.
  \item \textbf{Simultaneous} ($E_{sim}$): All slots are used at the same time. This is a common mode for many inherence-based techniques~\cite{Authentication.Techniques}.
  \item \textbf{Conditional} ($E_{cond}$): Exactly one slot is used. The choice depends on a condition, such as the user's environment or choice. Adaptive authentication uses this mode~\cite{Arias-Cabarcos.2019}.
\end{itemize}

%\noindent To define a slot order, we establish the convention that assumes slots are always ordered left to right.

\paragraph{Nesting:}
The above modes can model all basic compositions of authenticators that use single employments. However, employments can also be nested. For example, the following models a valid composition: ``authenticate first with a \password, then with either a \fingerprint\ or a \pin''.

We can achieve this by extending the mode concept with recursion. For each mode, we allow its arguments to be slots or, recursively, other multi employments, which we refer to as \emph{sub-employments}. The modes then apply to arguments instead of slots: sequential and simultaneous employments use all arguments, and a conditional employment uses exactly one argument, which may itself be a sub-employment comprising several slots. We can model the above with a sequential employment that specifies a slot in its first position, followed by a conditional sub-employment with two slots.

\paragraph{Notation:}
We introduce the following formal notation for these concepts: 
\begin{itemize}
  \item We denote a slot by $\varepsilon$ and an arbitrary employment by $E$. A single employment, i.e., an employment with exactly one slot, is $E(\varepsilon)$.
  \item Let $X_1,\dots,X_n$ with $n\ge 2$ be slots or employments and $\mathit{mode}\in\{seq,sim,cond\}$. Then $E_{\mathit{mode}}(X_1,\dots,X_n)$ is an employment. Those $X_i$ that are employments are its sub-employments. Its slots are the slots among the $X_i$ and the slots of the sub-employments, ordered from left to right.
  \item The cardinality $|E|$ is the number of slots $E$ specifies: $|E(\varepsilon)| = 1$, and the cardinality of $E_{\mathit{mode}}(X_1,\dots,X_n)$ is the sum over its arguments, where each slot counts as one and each sub-employment $X_i$ counts as $|X_i|$.
\end{itemize}

\paragraph{Examples:}
\begin{enumerate}
    \item $E = E_{seq}(\varepsilon, \varepsilon)$ is a sequential employment with two slots, i.e., $|E| = 2$.
    \item The aforementioned composition is modeled by $E = E_{seq}\big(\varepsilon, E_{cond}(\varepsilon, \varepsilon)\big)$. $E$ has one slot and one sub-employment $E_{cond}$ with 2 slots, i.e., $|E|=3$.
\end{enumerate}

\subsection{Authentication Technique}\label{sec:technique}

An employment specifies slots. These are independent of authenticators, i.e., each slot can be assigned an arbitrary authenticator. The \emph{authentication technique} defines the concrete assignment of which authenticator is assigned to which slot. The authenticators are then used in the mode and order that the employment specifies. %We say that a technique \emph{conforms to} an employment. This holds by construction: the authenticators of a technique fill exactly the slots of the employment.

For example, consider the employment $E_{seq}(\varepsilon,\varepsilon)$, specifying the sequential use of two authenticators. Assigning a \password\ to the first slot and a \fingerprint\ to the second slot yields the technique ``first use a \password, then scan a \fingerprint''. Assigning a \pin\ and an \iris\ \ scan yields a different technique, ``first \pin, then \iris'', conforming to the same employment.

\paragraph{Notation:} $A$ denotes the set of authenticators, $a \in A$ an authenticator. Formally, an authentication technique is a pair $T = (E, (a_1,\dots,a_k))$, where $E$ is an employment with cardinality $k$ and $a_i \in A$ is the authenticator assigned to the $i$-th slot of $E$.

While this notation is formally correct, as it defines which authenticator is assigned to which slot of an employment, it is not very readable. Thus, we introduce the \emph{substitution form} of an authentication technique. In substitution form, an authentication technique $T$ is expressed as a term of an employment $E$ in which the $i$-th slot $\varepsilon$ is substituted by the $i$-th authenticator $a $ that $T$ assigns. For example, for $E = E_{seq}(\varepsilon,\varepsilon)$ and $T = \big(E, (\text{\password}, \text{\fingerprint})\big)$, we write

\[
  T := E_{seq}(\text{\password}, \text{\fingerprint}),
\]
and read ``technique $T$ authenticates a user using a \password\ and a \fingerprint\ in sequential order''. If the same authenticator is assigned to several slots, we mark the further occurrences with primes, e.g., $T := E_{seq}(\text{\password, \password}')$: both authenticators are the same, but can be instantiated by a different concrete authenticator.

\paragraph{Examples:}
\begin{enumerate}
  \item $T := E_{seq}(\text{\password}, \text{\fingerprint})$ conforms to a sequential employment $E_{seq}(\varepsilon,\varepsilon)$, assigning a \password\ first, then a \fingerprint.% It conforms to $E_{seq}(\varepsilon,\varepsilon)$.
  \item $T := E_{cond}(\text{\password}, \text{\fingerprint})$ conforms to conditional employment $E_{cond}(\varepsilon, \varepsilon)$, assigning a \password\ for the first condition and a \fingerprint\ for the second.
  \item Consider the following setting: In a known environment, a user must first use a \password\ and then either scan a \fingerprint\ or use a \pin. In an unknown environment, the user must first use a \password, then scan their \iris, and then either type a \pin\ or provide an \otp. The following employment $E$ models this:
        \[
          E = E_{cond}\big(E_{seq}(\varepsilon, E_{cond}(\varepsilon,\varepsilon)),\;
                             E_{seq}(\varepsilon, \varepsilon, E_{cond}(\varepsilon,\varepsilon))\big),
        \]
        and the following technique $T$ conforms to $E$:
        \[
          T := E_{cond}\Big(E_{seq}\big(\text{\password}, E_{cond}(\text{\fingerprint},\text{\pin})),\;
                              E_{seq}(\text{\password}', \text{\iris}, E_{cond}(\text{\pin', \otp})\big)\Big)
        \]
\end{enumerate}

\subsection{Authentication Solution}
The authentication technique reflects the conceptual approach for how authentication is performed: which authenticators are used, and in which mode and order. Its technical realization is an \emph{authentication solution}. We align its definition with that of security solutions by NIST~\cite{NIST.800-172} as the key design, architectural, and implementation choices made to realize authentication. One such key design and architectural decision is what authentication technique an authentication solution \emph{implements}.

For example, consider the two authentication techniques $T_1:=E_{seq}(\text{\password, \fingerprint})$ and $T_2 := E_{seq}(\text{\password , \otp })$. The decision to implement $T_1$ rather than $T_2$ is a central design decision that significantly affects the overall authentication design and architecture: a secure password-based authentication solution must be designed~\cite{Distrinet}; users must have access to a fingerprint sensor; if necessary, appropriate sensors must be purchased, deployed, and maintained, etc.~\cite{Maltoni.2022}. In contrast, choosing $T_2$ requires designing secure password-based authentication, along with a proper client-side or server-side OTP mechanism~\cite{Berenjestanaki.2019}, and secret management and enrollment processes that differ from those required by a fingerprint solution~\cite{Huseynov.2017}. In conclusion, the authentication technique reflects the conceptual solution, while the design decisions for password-based, fingerprint-based, and OTP-based authentication are made at the authentication solution level.

\paragraph{Notation:}
A solution implements a technique together with authenticator constraints and solution-specific design and implementation decisions. An example of a constraint is a minimum password length. Constraints are not within the scope of this paper and remain for future work. For design and implementation decisions, we omit a formal notation, as they are open-ended. Instead, we capture their consequences at the metric level. This will be described in detail in Section~\ref{sec:solution-quality}. We therefore only name the technique a solution $S$ implements, and write ``$S$ implements $T$''.

\paragraph{Examples:}
\begin{enumerate}
  \item A single-factor authentication solution that uses a password for authentication implements $T := E(\text{\password})$.
  \item An MFA solution that uses a password as first factor and a client-side OTP as second factor implements $T := E_{seq}(\text{\password, \otp})$. The solution's OTP mechanism is a design decision; we do not model it in this notation, but we capture its consequences at the metric level.
  \item Two solutions can implement the same technique. For $T := E_{seq}(\text{\password, \fingerprint})$, a solution $S_1$ implements $T$ as an Android app and a solution $S_2$ as a desktop app. They are different solutions that implement the same technique because they realize the technique on different platforms.
\end{enumerate}

\section{Quality Relationships}\label{sec:quality}
As shown in Figure~\ref{fig:concepts}, authenticators, authentication techniques, and authentication solutions have qualitative properties. These properties propagate from the authenticator, which resembles the atomic concept, to authentication techniques and then to authentication solutions. This is indicated by the \emph{depends on} relationships. We illustrate this using the quality factor \textbf{usability}, focusing on a specific subfactor: the \textbf{Ease of Use (EoU)}. It expresses the degree to which it is easy for a user to successfully authenticate with an authentication solution.

\subsection{Authenticator Quality}
To evaluate a solution's EoU, we must first assess the authenticator's EoU, since the authenticator is what a user uses to authenticate. Consequently, part of a solution's EoU depends on the authenticators it employs. 

For an authenticator, we define the EoU as ``the degree of how easy an authenticator can be inherently used by a human for authentication purposes.'' For illustration, we define a simple counting metric for EoU as the number of user interactions with an input device that an authenticator inherently requires to successfully authenticate a user. We divide 1 by this number to obtain a value on the scale $[0,1]$, and interpret the metric such that the higher the value, the better the EoU:

\begin{align*}
    \mathrm{EoU}(a) = \frac{1}{\text{number of user interactions}}
\end{align*}

\paragraph{Examples:}
\begin{enumerate}
    \item $\mathrm{EoU}(\text{\fingerprint}) = 1$, as it inherently enables authentication through a single touch.
    \item $\mathrm{EoU}(\text{\password}) = 1/\ell_{\text{\password}}$, since without tool support such as key management tools, a \password\ requires at least as many user interactions with an input device as the number of characters it consists of.
\end{enumerate}

\subsection{Authentication Technique Quality}

An authentication technique influences a solution’s EoU through the authenticators and their employment. We define a technique’s EoU as ``the degree to which all authenticators used can be easily utilized by a human for authentication purposes.'' The fundamental quality \emph{depends on} the authenticators. The employment \emph{determines} the technique's quality, since how the authenticators are used together affects the overall quality. To evaluate the technique’s quality, we must therefore examine how each employment influences that quality.

\paragraph{Single employments:} The EoU of a single employment is simply that of the authenticator employed:
\[
  \mathrm{EoU}\big(E(a)\big) = \mathrm{EoU}(a).
\]

\paragraph{Sequential and simultaneous employments:} 
The user must use all authenticators either sequentially or simultaneously. Since EoU counts interactions, not time effort, the number of user interactions adds up in both modes. As we divide 1 by the number of user interactions, we sum the reciprocals. A slot filled with an authenticator $a$ has the EoU of $a$. For slots or sub-employments $X_1,\dots,X_n$:
   \[
     \mathrm{EoU}(E_{seq}\big(X_1,\dots,X_n)\big) = \mathrm{EoU}\big(E_{sim}(X_1,\dots,X_n)\big) = \frac{1}{\sum_{i=1}^{n} 1/\mathrm{EoU}(X_i)}
   \]

This ensures that the result is always equal to or lower than the EoU of the least usable sub-employment; adding an authenticator can only worsen a technique's EoU. This aligns with the common perception that MFA negatively affects usability~\cite {Klivan.2023}.

\paragraph{Conditional employments:} The user uses one sub-employment. Thus, the EoU of a conditional employment is that of the sub-employment used. For slots or sub-employments $X_1,\dots,X_n$: 
\[
  \mathrm{EoU}\big(E_{cond}(X_1,\dots,X_n)\big) = \mathrm{EoU}(X_i),
  \quad \text{if } X_i \text{ is used}, \; i \in \{1,\dots,n\}.
\]

\paragraph{Examples:}
\begin{enumerate}
  \item For a technique $T := E(\text{\fingerprint})$, it is $\mathrm{EoU}(T) = 1$.
  \item For a technique $T := E_{seq}(\text{\password, \fingerprint})$, it is $
          \mathrm{EoU}(T)
          = \frac{1}{1/\mathrm{EoU}(\text{\password}) + 1/\mathrm{EoU}(\text{\fingerprint})}
          = \frac{1}{\ell_{\text{\password}}+1}.
        $
        
        When $T$ is used with a concrete \password\ with length $\ell_{\text{\password}}=9$, it yields $\mathrm{EoU}(T) = 1/10 = 0.1$. When $T$ is used with a concrete \password\ with length $\ell_{\text{\password}}=19$, it yields $\mathrm{EoU}(T) = 1/20 = 0.05$. Thus, its EoU decreases with longer passwords, as expected.
  \item For a technique $T := E_{cond}(\text{\password, \fingerprint})$, it is
        \[
          \mathrm{EoU}(T) =
          \begin{cases}
            1/\ell_{\text{\password}}, & \text{if } \text{\password} \text{ is used},\\
            1,           & \text{if } \text{\fingerprint} \text{ is used}.
          \end{cases}
        \]
  \item For the nested technique $T$ from Example 3 in Section~\ref{sec:technique},
        \[
          T := E_{cond}\Big(E_{seq}\big(\text{\password}, E_{cond}(\text{\fingerprint},\text{\pin})),\;
                              E_{seq}(\text{\password}', \text{\iris}, E_{cond}(\text{\pin}',\text{\otp})\big)\Big).
        \]
        We evaluate this bottom-up. Assume $\mathrm{EoU}(\text{\fingerprint})=\mathrm{EoU}(\text{\iris})=1$, as both are biometric authenticators that allow authentication with a single interaction~\cite{Authentication.Patterns}.
        
        Moreover, assume
        $\ell_{pw}=9$, $\ell_{pw'}=19$, $\ell_{pin}=\ell_{pin'}=4$, and $\ell_{otp}=6$, treating PINs as short passwords and OTPs as single-use passwords~\cite{Authentication.Patterns}.
        Then, the innermost conditional employments yield:
        \[
          \mathrm{EoU}\big(E_{cond}(\text{\fingerprint},\text{\pin})\big) =
          \begin{cases} 1, & \text{if } \text{\fingerprint} \text{ is used},\\
                        1/4, & \text{if } \text{\pin} \text{ is used},\end{cases}
          \quad
          \mathrm{EoU}\big(E_{cond}(\text{\pin}',\text{\otp})\big) =
          \begin{cases} 1/4, & \text{if } \text{\pin}' \text{ is used},\\
                        1/6, & \text{if } \text{\otp} \text{ is used}.\end{cases}
        \]
        Next, the sequential employments sum the interactions, and the
        outermost conditional employment selects one of them:
        \[
          \mathrm{EoU}(T) =
          \begin{cases}
            1/(9+1)    = 0.1,        & \text{if } \text{\password}, \text{\fingerprint} \text{ are used},\\
            1/(9+4)    \approx 0.077, & \text{if } \text{\password}, \text{\pin} \text{ are used},\\
            1/(19+1+4) \approx 0.042, & \text{if } \text{\password}', \text{\iris}, \text{\pin}' \text{ are used},\\
            1/(19+1+6) \approx 0.038, & \text{if } \text{\password}', \text{\iris}, \text{\otp} \text{ are used}.
          \end{cases}
        \]
        Thus, $T$ has four values, one for each combination of selected
        alternatives. The best case is more than twice as easy to use as the
        worst case, which a single number for the technique would hide.
\end{enumerate}

\subsection{Authentication Solution Quality}\label{sec:solution-quality}

On the solution level, we define the EoU as ``the degree to which the authentication solution as a whole can be easily used by a human for authentication purposes''. It is essentially determined by the EoU of the technique $T$ it implements, but the solution's design and implementation decisions still affect the final value. Consider the technique $T := E_{seq}(\text{\password, \fingerprint})$ and two solutions implementing it: \textit{MFA-1} for an Android or iOS app; \textit{MFA-2} for a desktop app. For illustration, we assume that autofill and pasting of passwords are disabled.

Although both solutions implement $T$, their EoU differs. For MFA-1, the fingerprint sensor is integrated into the smartphone, so both authenticators are used on the same device. However, entering a password on a touchscreen is cumbersome and thus reduces usability~\cite{Schaub.2012}. For MFA-2, the \password\ is entered on a standard keyboard, but most keyboards lack a fingerprint sensor, so the user must switch to another device. This affects the quality, as context switches reduce usability~\cite{Lennartsson.2021}.

To account for such solution-specific factors, we introduce two variables analogous to effort multipliers in estimation models such as COCOMO~\cite{Boehm.1981}:

\begin{description}
  \item[input factor $f_{in}$:] reflects factors that simplify or complicate how authenticators are entered. Without such factors, $f_{in}=1$; simplifying factors give $f_{in}>1$, complicating factors $f_{in}<1$.
  \item[context switch disruption ($f_{cs}$):] reflects how much usability is impaired by context switches. We rank them by: $f_{cs}=0$ no context switch required; $f_{cs}=0.1$ user must switch the application but not the device; $f_{cs}=0.2$ if the user must switch devices.
\end{description}
The EoU of a solution $S$ implementing $T$ can then be calculated by the formula on the scale $[0,1]$:
\[
  \mathrm{EoU}(S) = \min\big(1,\; \mathrm{EoU}(T) \cdot f_{in} \cdot (1- f_{cs})\big).
\]
%Since both factors only scale the EoU of the technique, the result stays on the scale $[0,1]$.

\paragraph{Examples:} Consider $T := E_{seq}(\text{\password, \fingerprint})$, with $\mathrm{EoU}(T) = \frac{1}{\ell_{\text{\password}}+1}$, and two passwords $\text{\password}_A$ with $\ell_{\text{\password}_A}=9$, $\text{\password}_B$ with $\ell_{\text{\password}_B}=19$.
\begin{enumerate}
  \item For MFA-1, we choose $f_{in}=0.9$, due to the complicated password input, and $f_{cs}=0$, as there is no context switch. It is $\mathrm{EoU}(\text{MFA-1}) = \mathrm{EoU}(T) \cdot 0.9$. For $\text{\password}_A$ it is $0.09$; for $\text{\password}_B$ it is $0.045$.
  \item For MFA-2, we choose $f_{in}=1$, as no input is complicated, and $f_{cs}=0.2$, as the user must change devices. It is $\mathrm{EoU}(\text{MFA-2}) = \mathrm{EoU}(T) \cdot 0.8$. For $\text{\password}_A$ this yields $0.08$; for $\text{\password}_B$ it is $0.04$.
\end{enumerate}

Given these illustrative metrics, MFA-1 has a better EoU than MFA-2. Note that this conclusion depends on the chosen values of $f_{in}$ and $f_{cs}$.

\paragraph{Summary:} We showed by example that each first-order concept in the \textsc{Aqua} framework directly relates to quality, and that this quality propagates from the authenticator to the technique that employs it to the solution it implements: An authenticator affects a solution's quality through its inherent properties; the technique affects it through the aggregation rules of the employment it conforms to; the solution affects it through design decisions made at the solution level. Lastly, the EoU example also shows that subfactors can be shared across concepts, with slightly different meanings. However, the sets of quality factors and subfactors are not equal. For example, privacy and its subfactors are defined at the solution level, as these qualities are affected by how a solution processes data.

\section{Demonstration: Creating Quality Models with \textsc{Aqua}}\label{sec:demo}
\begin{figure*}[!ht]
    \begin{center}
        \includegraphics[width=\textwidth]{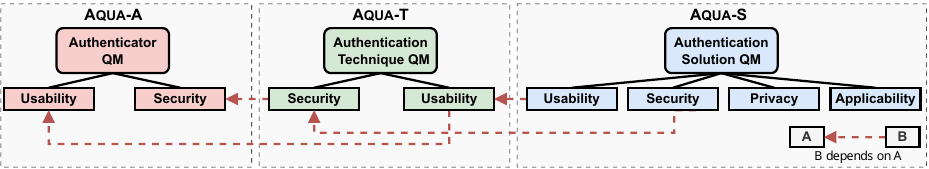}
    \caption{The three \textsc{Aqua} models including their quality factors and the dependencies between them.}
    \label{fig:aqua-models}
    \end{center}
\end{figure*}

To demonstrate the \textsc{Aqua} framework's modeling capabilities, we developed initial, non-exhaustive quality models for authenticators, authentication techniques, and authentication solutions. To this end, we iteratively analyzed the qualitative properties of each concept, starting with the authenticator and ending with the authentication solution. We sought to describe each concept using a set of comprehensible quality properties specific to that concept. They were documented and structured based on the ISO SQuaRE model \cite{ISO.25000} into quality factors and subfactors. In this course, we focused on usability and security, given the extensive body of knowledge in these areas~\cite{Bonneau.2012}. When defining the subfactors, we consistently verified whether each subfactor was inherent to the concept or to another.

The results are three interrelated quality models: \textsc{Aqua-A} for authenticators, \textsc{Aqua-T} for techniques, and \textsc{Aqua-S} for solutions. All models are definitional quality models, i.e., they do not include metrics. Moreover, they are not exhaustive and should be seen as clearly preliminary results. However, they illustrate the structure of quality models built in \textsc{Aqua} and the quality dependencies between them.

\begin{figure}[!t]
    \begin{center}
        \includegraphics[width=\textwidth]{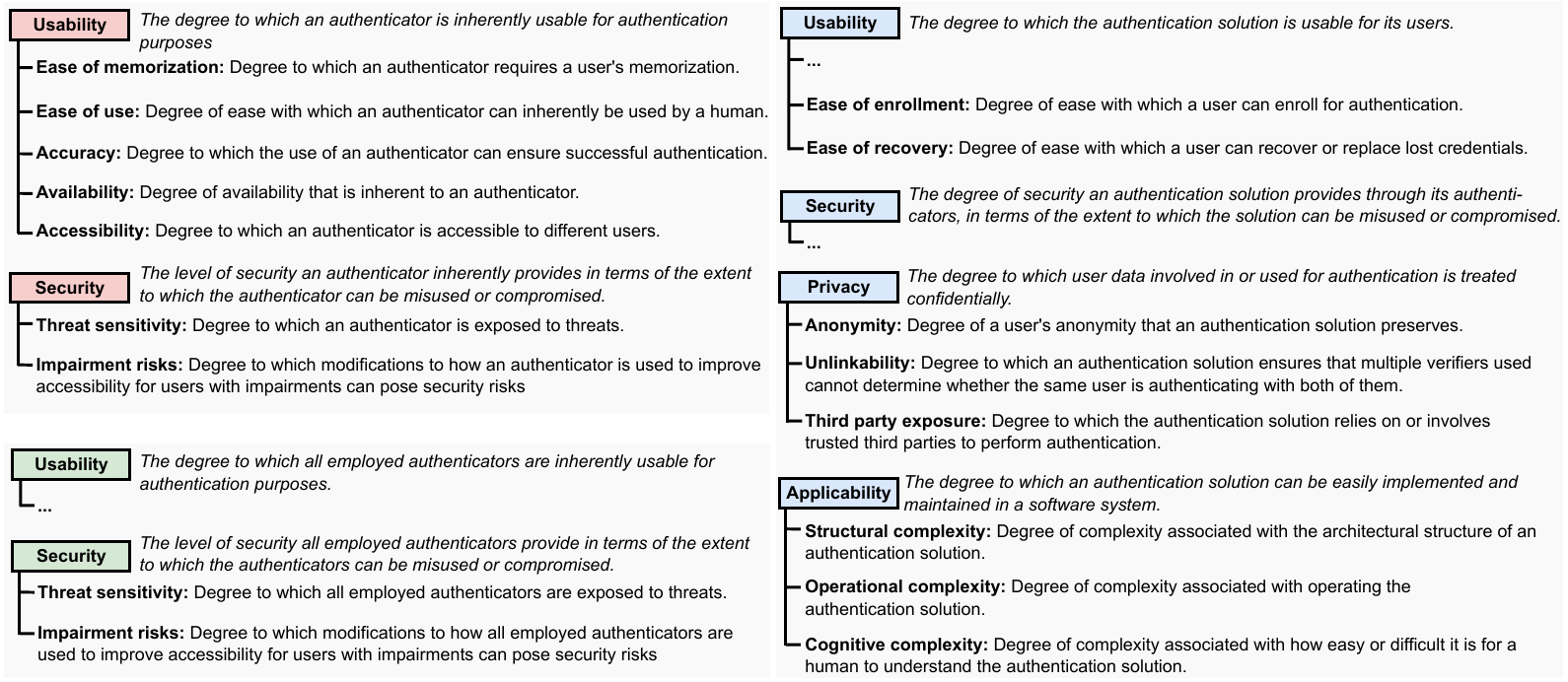}
    \caption{Subfactors of the \textsc{Aqua} quality models. Of the dependent quality factors and subfactors, only \textsc{Aqua-T}'s Security subfactors are depicted to illustrate the similarity among dependent factors.}
    \label{fig:aqua-model-details}
    \end{center}
\end{figure}

Figure~\ref{fig:aqua-models} shows the three quality models at the quality factor level and the dependent quality factors. Figure~\ref{fig:aqua-model-details} depicts the subfactors. The colors indicate which quality model the factors and subfactors belong to: red for \textsc{Aqua-A}; green for \textsc{Aqua-T}; blue for \textsc{Aqua-S}. As the dependent factors and subfactors read similarly, we do not depict most of them and share only the subfactors of \textsc{Aqua-T}'s security factor for illustration. The complete documentation of all \textsc{Aqua} models, including reasoning and an example for each subfactor, can be found on the \textsc{Aqua} webpage\footnote{\url{https://scam-research-project.gitlab.io/aqua/}}.

\section{Conclusions}\label{sec:conclusions}

This paper presents the \textsc{Aqua} framework. \textsc{Aqua} is a modular framework for modeling authentication quality. It decomposes authentication into four interrelated concepts: the authenticators; composition schemes for them through authenticator employments; their concrete employments through authentication techniques; their implementation through authentication solutions. This decomposition forms the basis for a modular approach to modeling authentication quality and is presented using a formal notation we developed, along with comprehensive examples that illustrate their relationships to quality. To demonstrate, we developed initial definitional quality models for authenticators, authentication techniques, and authentication solutions, and deployed them as an interactive web application.

\subsection{Limitations}

This study is subject to certain limitations. It elaborates on the inherent quality properties of authenticators---that is, how authenticators, by their very nature, influence the overall quality of an authentication solution. To illustrate this, we accepted a clear limitation: we did not allow users to copy and paste or autofill passwords. However, this limitation was necessary to highlight the qualitative nature underlying the authenticators. Our interpretation of an authenticator's ease of use, i.e., counting the number of required user interactions with an input device to use an authenticator for authentication, implies for passwords that they require $n$ user inputs, one per character. This is a property inherent to a password. However, tools such as password managers mitigate this drawback by letting users copy and paste passwords. Using a password manager does, however, not mean that the inherent property of the password is thereby rendered obsolete. Rather, password managers owe their very existence to the fact that passwords are, among other things, poorly usable for authentication.

Next, because of their qualitative nature, both the decomposition underlying the \textsc{Aqua} framework and the \textsc{Aqua} models may have been influenced by subjectivity. We tried to minimize this threat by making decisions only after the authors agreed. Moreover, we verified \textsc{Aqua}'s internal consistency and expressiveness by formalizing the concepts.

Furthermore, although the three \textsc{Aqua} models are substantial results on their own, they are limited to a small number of factors and subfactors. Given the project's early stage, this limitation is acceptable. Moreover, we developed them for demonstration purposes and therefore do not claim they are comprehensive. Expanding these factors is part of future work. Lastly, the EoU metrics are purely illustrative, based on our experience, and have not yet been evaluated.

\subsection{Future Work}

\textsc{Aqua} opens up many important future directions. As mentioned in the limitations, we plan to extend the \textsc{Aqua} models with additional quality factors and subfactors. We also plan to extend \textsc{Aqua} with assessment and prediction models; the metrics we used for demonstration already capture some of our ideas. Such models will also support a comprehensive evaluation, which we consider another central direction for future work.

Next, we consider \textsc{Aqua} as a foundation for effective automated quality prediction and assessment in authentication solutions. Future studies could investigate how to use \textsc{Aqua} for these purposes. Moreover, future work can focus on establishing quality profiles with \textsc{Aqua}. The \textsc{Aqua} models we designed allow us to define quality profiles for authenticators and authentication techniques. For example, separate profiles for generated and self-chosen passwords can be defined, where the generated-password profile ranks memorability lower but threat sensitivity higher than the self-chosen profile.

%Lastly, we see a valuable addition to \textsc{Aqua} by developing a cost model for authentication solutions. In our efforts to develop the \textsc{Aqua} models, we identified clear indicators for predicting the costs that arise from specific authentication solutions. For example, costs are implied by the sensor hardware required to use inherence-based authenticators, or the management of certificates and public key infrastructures.

\section*{Declaration on Generative AI}
The authors used: 1) Claude to test the formal notation for errors and inconsistencies; 2) Claude, Grammarly, and DeepL for grammar and spelling checks, and to improve writing style. After use, the authors reviewed and edited the content as needed. They take full responsibility for the paper’s content.

\bibliographystyle{unsrt}  
\bibliography{Bibliography}

@article{Furnell.2022,
title = {Accessible authentication: Assessing the applicability for users with disabilities},
journal = {Computers \& Security},
volume = {113},
pages = {102561},
year = {2022},
issn = {0167-4048},
doi = {https://doi.org/10.1016/j.cose.2021.102561},
author = {Steven Furnell and Kirsi Helkala and Naomi Woods}
}

@misc{OWASP.Top10,
  author       = {{The OWASP Foundation}},
  title        = {{OWASP Top 10:2025}},
  organization = {{Open Worldwide Application Security Project (OWASP)}},
  url = {https://owasp.org/Top10/2025/},
  year         = {2026},
  note         = {last accessed: 2026-10-01}
}

@article{Peffers.2007,
 author = {Peffers, Ken and Tuunanen, Tuure and Rothenberger, Marcus A. and Chatterjee, Samir},
 year = {2007},
 title = {A Design Science Research Methodology for Information Systems Research},
 volume = {24},
 number = {3},
 issn = {0742-1222},
 journal = {Journal of Management Information Systems},
 doi = {10.2753/MIS0742-1222240302}
}

@inproceedings{Bonneau.2012,
author = {Bonneau, Joseph and Herley, Cormac and Oorschot, Paul C. van and Stajano, Frank},
title = {The Quest to Replace Passwords: A Framework for Comparative Evaluation of Web Authentication Schemes},
year = {2012},
isbn = {9780769546810},
publisher = {IEEE Computer Society},
address = {USA},
doi = {10.1109/SP.2012.44},
booktitle = {Proceedings of the 2012 IEEE Symposium on Security and Privacy},
pages = {553–567},
numpages = {15},
series = {SP '12}
}

@book{Boehm.1981,
  author    = {Boehm, Barry W.},
  title     = {Software Engineering Economics},
  publisher = {Prentice-Hall},
  address   = {Englewood Cliffs, NJ},
  year      = {1981},
  isbn      = {0-13-822122-7}
}

@article{Zafar.2015,
author = {Zafar, Saad and Mehboob, Misbah and Naveed, Asma and Malik, Bushra},
title = {Security quality model: an extension of Dromey's model},
year = {2015},
issue_date = {March     2015},
publisher = {Kluwer Academic Publishers},
address = {USA},
volume = {23},
number = {1},
issn = {0963-9314},
doi = {10.1007/s11219-013-9223-1},
journal = {Software Quality Journal},
month = mar,
pages = {29–54},
numpages = {26}
}

@inproceedings{Xu.2013,
author = {Xu, Haiyun and Heijmans, Jeroen and Visser, Joost},
title = {A Practical Model for Rating Software Security},
year = {2013},
isbn = {9781479929252},
publisher = {IEEE Computer Society},
address = {USA},
doi = {10.1109/SERE-C.2013.11},
booktitle = {Proceedings of the 2013 IEEE Seventh International Conference on Software Security and Reliability Companion},
pages = {231–232},
numpages = {2},
series = {SERE-C '13}
}

@inproceedings{Aziz.2018,
  author={Aziz, M. Nasrul and Sapta, Irit Maulana and Rochimah, Siti},
  booktitle={2018 Electrical Power, Electronics, Communications, Controls and Informatics Seminar (EECCIS)}, 
  title={Security Characteristic Evaluation Based on ISO/IEC 25023 Quality Model, Case Study: Laboratory Management Information System}, 
  year={2018},
  volume={},
  number={}
}

@techreport{NIST.800-172,
 author = {Pillitteri, Victoria and Ross, Ron},
 title = {Enhanced Security Requirements for Protecting Controlled Unclassified Information: NIST Special Publication 800-172r3},
 number = {800-172r3},
 series = {NIST Special Publication},
 editor = {{U.S. Department of Commerce}},
 institution = {{National Institute of Standards and Technology}},
 doi = {10.6028/NIST.SP.800-172r3},
 year = 2026
}

@InProceedings{Chenchev.2021,
author="Chenchev, Ivaylo
and Aleksieva-Petrova, Adelina
and Petrov, Milen",
editor="Arai, Kohei",
title="Authentication Mechanisms and Classification: A Literature Survey",
booktitle="Intelligent Computing",
year="2021",
publisher="Springer International Publishing",
address="Cham",
pages="1051--1070",
isbn="978-3-030-80129-8"
}

@book{Todorov.2007,
 author = {Todorov, Dobromir},
 year = {2007},
 title = {Mechanics of User Identification and Authentication: Fundamentals of Identity Management},
 edition = {1},
 publisher = {{Auerbach Publications}},
 isbn = {9781040169469}
}

@misc{Distrinet,
  author       = {{DistriNet}},
  title        = {{Security Pattern Catalog}},
  organization = {{imec-DistriNet research group, KU Leuven}},
  url = {https://securitypatterns.distrinet-research.be/},
  year         = {2026},
  note         = {last accessed: 2026-08-01}
}

@inproceedings{Klivan.2023,
author = {Klivan, Sabrina and H\"{o}ltervennhoff, Sandra and Huaman, Nicolas and Krause, Alexander and Simko, Lucy and Acar, Yasemin and Fahl, Sascha},
title = {"We've Disabled MFA for You": An Evaluation of the Security and Usability of Multi-Factor Authentication Recovery Deployments},
year = {2023},
isbn = {9798400700507},
publisher = {Association for Computing Machinery},
address = {New York, NY, USA},
doi = {10.1145/3576915.3623180},
booktitle = {Proceedings of the 2023 ACM SIGSAC Conference on Computer and Communications Security},
pages = {3138–3152},
numpages = {15},
location = {Copenhagen, Denmark},
series = {CCS '23}
}

@Article{Matzen.2025,
AUTHOR = {Matzen, Alexander and Rüffer, Artur and Byllemos, Marcus and Heine, Oliver and Papaioannou, Maria and Choudhary, Gaurav and Dragoni, Nicola},
TITLE = {Challenges and Potential Improvements for Passkey Adoption—A Literature Review with a User-Centric Perspective},
JOURNAL = {Applied Sciences},
VOLUME = {15},
YEAR = {2025},
NUMBER = {8},
ARTICLE-NUMBER = {4414},
ISSN = {2076-3417},
DOI = {10.3390/app15084414}
}

@article{Mahjabin.2017,
author = {Mahjabin, Tasnuva and Xiao, Yang and Sun, Guang and Jiang, Wangdong},
title ={A survey of distributed denial-of-service attack, prevention, and mitigation techniques},
journal = {International Journal of Distributed Sensor Networks},
volume = {13},
number = {12},
pages = {1550147717741463},
year = {2017},
doi = {10.1177/1550147717741463},
}

@techreport{NIST.800-63-4,
 author = {{David Temoshok} and {Diana Proud-Madruga} and {Yee-Yin Choong} and {Ryan Galluzzo} and {Sarbari Gupta} and {Connie LaSalle} and {Naomi Lefkovitz} and {Andrew Regenscheid}},
 year = {2025},
 title = {Digital Identity Guidelines: NIST Special Publication},
 number = {SP 800-63-4},
 editor = {{U.S. Department of Commerce}},
 institution = {{National Institute of Standards and Technology (NIST)}},
 doi = {10.6028/NIST.SP.800-63-4}
}

@techreport{NIST.800-37,
 author = {{Joint Task Force}},
 year = {2018},
 title = {Risk Management Framework for Information Systems and Organizations: A System Life Cycle Approach for Security and Privacy: NIST Special Publication 800-37 Revision 2},
 number = {NIST SP 800-37 Rev. 2},
 series = {Federal Information Processing Standards Publication},
 editor = {{U.S. Department of Commerce}},
 institution = {{National Institute of Standards and Technology}},
}

@inproceedings{Neil.2023,
 author = {Neil, Lorenzo and Haney, Julie M. and Buchanan, Kerrianne and Healy, Charlotte},
 title = {Analyzing Cybersecurity Definitions for Non-experts},
 pages = {391--404},
 publisher = {{Springer Nature Switzerland}},
 isbn = {978-3-031-38530-8},
 editor = {{Furnell, Steven and Clarke, Nathan}},
 booktitle = {Human Aspects of Information Security and Assurance},
 year = {2023},
 address = {Cham}
}

@article{Arias-Cabarcos.2019,
author = {Arias-Cabarcos, Patricia and Krupitzer, Christian and Becker, Christian},
title = {A Survey on Adaptive Authentication},
year = {2019},
issue_date = {July 2020},
publisher = {Association for Computing Machinery},
address = {New York, NY, USA},
volume = {52},
number = {4},
issn = {0360-0300},
url = {https://doi.org/10.1145/3336117},
doi = {10.1145/3336117},
journal = {ACM Comput. Surv.},
month = sep,
articleno = {80},
numpages = {30}
}

@inproceedings{Authentication.Techniques,
author={Alex R. Mattukat and Vincent Schmandt and Timo Langstrof and Michael Zerbe and Horst Lichter},
title={A Faceted Classification of Authenticator-Centric Authentication Techniques},
booktitle={Proceedings of the 21st International Conference on Evaluation of Novel Approaches to Software Engineering - Volume 1: ENASE},
year={2026},
pages={568-578},
publisher={SciTePress},
organization={INSTICC},
doi={10.5220/0014982200004015},
isbn={978-989-758-828-0},
issn={2184-4895},
}

@article{Ometov.2018,
AUTHOR = {Ometov, Aleksandr and Bezzateev, Sergey and Mäkitalo, Niko and Andreev, Sergey and Mikkonen, Tommi and Koucheryavy, Yevgeni},
TITLE = {Multi-Factor Authentication: A Survey},
JOURNAL = {Cryptography},
VOLUME = {2},
YEAR = {2018},
NUMBER = {1},
ARTICLE-NUMBER = {1},
ISSN = {2410-387X},
DOI = {10.3390/cryptography2010001}
}

@ARTICLE{Authentication.Patterns,
       author = {{Mattukat}, Alex R. and {Lichter}, Horst},
        title = "{A Catalog of User Authentication Patterns}",
      journal = {arXiv e-prints},
         year = 2026,
        month = aug,
          eid = {arXiv:2608.26955},
        pages = {arXiv:2608.26955},
          doi = {10.48550/arXiv.2608.26955},
 primaryClass = {cs.CR},
       adsurl = {https://ui.adsabs.harvard.edu/abs/2026arXiv260826955M}
}

@inproceedings{Berenjestanaki.2019,
author = {Berenjestanaki, Md. Hajian and Conti, Mauro and Gangwal, Ankit},
title = {On the Exploitation of Online SMS Receiving Services to Forge ID Verification},
year = {2019},
isbn = {9781450371643},
publisher = {Association for Computing Machinery},
address = {New York, NY, USA},
doi = {10.1145/3339252.3339276},
booktitle = {Proceedings of the 14th International Conference on Availability, Reliability and Security},
articleno = {22},
numpages = {5},
location = {Canterbury, CA, United Kingdom},
series = {ARES '19}
}

@incollection{Huseynov.2017,
title = {Chapter 50 - Context-Aware Multifactor Authentication Survey},
editor = {John R. Vacca},
booktitle = {Computer and Information Security Handbook (Third Edition)},
publisher = {Morgan Kaufmann},
edition = {Third Edition},
address = {Boston},
pages = {715-726},
year = {2017},
isbn = {978-0-12-803843-7},
doi = {https://doi.org/10.1016/B978-0-12-803843-7.00050-8},
author = {Huseynov, Emin and Seigneur, Jean-Marc}
}

@book{Maltoni.2022,
 author = {Maltoni, Davide and Maio, Dario and Jain, Anil K. and Feng, Jianjiang},
 year = {2022},
 title = {Handbook of Fingerprint Recognition},
 edition = {3},
 publisher = {{Springer Nature}},
 isbn = {978-3-030-83624-5}
}

@techreport{ISO.25000,
 year = {2023},
 title = {Systems and software engineering - Systems and software Quality Requirements and Evaluation (SQuaRE) - Product quality model},
 edition = {2},
 number = {ISO/IEC 25010},
 author = {{ISO/IEC 25010}},
 institution = {{International Organization for Standardization and International Electrotechnical Commission}},
 type = {Standard}
}

@inproceedings{Schaub.2012,
  title={Password entry usability and shoulder surfing susceptibility on different smartphone platforms},
  author={Schaub, Florian and Deyhle, Ruben and Weber, Michael},
  booktitle={Proceedings of the 11th international conference on mobile and ubiquitous multimedia},
  pages={1--10},
  year={2012}
}

@article{Lennartsson.2021,
    author = {Lennartsson, Markus and Kävrestad, Joakim and Nohlberg, Marcus},
    title = {Exploring the meaning of usable security – a literature review},
    journal = {Information and Computer Security},
    volume = {29},
    number = {4},
    pages = {647-663},
    year = {2021},
    issn = {2056-4961},
    doi = {10.1108/ICS-10-2020-0167}
}

\end{document}